\documentclass[11pt,a4paper,reqno]{amsart}
\usepackage{multirow}
\usepackage[centertags]{amsmath}
\usepackage{amsfonts}
\usepackage{amssymb}
\usepackage{amsthm}
\usepackage{newlfont}
\usepackage{a4}
\usepackage{enumerate}
\usepackage[pdftex]{graphicx,color}
\theoremstyle{definition}
\newtheorem{defn}{Definition}[section]
\newtheorem{thm}[defn]{Theorem}

\newtheorem{tvr}[defn]{Proposition}
\newtheorem{cor}[defn]{Corollary}
\theoremstyle{remark}
\newtheorem{example}{Example}[section]

\usepackage{bbm}

\newlength{\defbaselineskip}
\newcommand{\setlinespacing}[1]%
           {\setlength{\baselineskip}{#1 \defbaselineskip}}

\renewcommand{\i}{\mathrm{i}}
\newcommand{\map}{\rightarrow}

\newcommand{\q}{\quad}

\renewcommand{\epsilon}{\varepsilon}

\newcommand{\ep}{\varepsilon}
\newcommand{\la}{\lambda}
\newcommand{\al}{\alpha}
\newcommand{\om}{\omega}
\renewcommand{\rho}{\varrho}
\renewcommand{\phi}{\varphi}

\newcommand{\R}{{\mathbb{R}}}

\newcommand{\N}{{\mathbb N}}

\newcommand{\Com}{{\mathbb C}}

\newcommand{\Z}{\mathbb{Z}}

\newcommand{\set}[2]{\left\{#1 \, |\, #2 \right\}}
\newcommand{\setb}[2]{\left\{#1 \, \mid\, #2 \right\}}
\newcommand{\setm}[2]{\left\{#1 \, \big|\, #2 \right\}}
\newcommand{\abs}[1]{\left\vert#1\right\vert}
\newcommand{\wt}{\widetilde}

\newcommand{\sca}[2]{\langle #1,\, #2\rangle}
\newcommand{\comb}[2]{\begin{pmatrix}
     #1\\
     #2
  \end{pmatrix}}

\begin{document}

\title[Torus discretization]
{On Discretization of Tori of Compact Simple Lie Groups}

\author{Ji\v{r}\'{i} Hrivn\'{a}k$^{1,2}$}
\author{Ji\v{r}\'{i} Patera$^1$}

\date{\today}
\begin{abstract}\
Three types of numerical data are provided for simple Lie groups of any type and rank. This data is indispensable for Fourier-like expansions of multidimensional digital data into finite series of $C-$ or $S-$functions on the fundamental domain $F$ of the underlying Lie group $G$. Firstly, we consider the number $|F_M|$ of points in $F$ from the lattice $P^{\vee}_M$, which is the refinement of the dual weight lattice $P^{\vee}$ of $G$ by a positive integer $M$. Secondly, we find the lowest set $\Lambda_M$ of dominant weights, specifying the maximal set of $C-$ and $S-$functions that are pairwise orthogonal on the point set $F_M$. Finally, we describe an efficient algorithm for finding, on the maximal torus of $G$, the number of conjugate points to every point of $F_M$. Discrete $C-$ and $S-$transforms, together with their continuous interpolations, are presented in full generality. 

\end{abstract}\

\maketitle

\noindent
$^1$ Centre de recherches math\'ematiques,
         Universit\'e de Montr\'eal,
         C.~P.~6128 -- Centre ville,
         Montr\'eal, H3C\,3J7, Qu\'ebec, Canada; patera@crm.umontreal.ca\\
$^2$ Department of physics,
Faculty of nuclear sciences and physical engineering, Czech
Technical University, B\v{r}ehov\'a~7, 115 19 Prague 1, Czech
republic; jiri.hrivnak@fjfi.cvut.cz


\section{Introduction}

Processing of multidimensional digital data is the general motivation for this paper. However, our immediate objective is much more specific: recently introduced new Fourier like transforms, called $C-$, $S-$, and $E-$transforms \cite{P}, or simply orbit function transforms, on lattices of any dimension, symmetry and density, reviewed in \cite{KP1,KP2,KP3} (see also references therein), require certain technical data to be truly versatile and useful tools in all cases. Providing such data in full generality is the main goal of this paper. For the lowest dimension, $n=2$ and $3$, it was possible to obtain the data by simpler means, although, even then it presented a challenge of sorts \cite{PZ1,PZ2,PZ3,PA,NP}.

The theoretical background of this paper is the amazing uniformity, as to the type and rank, of the theory of simple Lie algebras over the complex number field \cite{Bour,H1,VO} and related finite Weyl groups \cite{H2,BB}, also referred to as crystallographic Coxeter groups. Discretization problems frequently call for properties of the maximal torus of the underlying compact simple Lie group \cite{MP3}.

Data provided in the paper can be used in many investigations. The focus is on three types of transforms defined on a well-known finite region $F$ of a real Euclidean space of dimension $n$, where $n$ is the rank of the underlying simple Lie group $G$. The region for digital data is the lattice fragment $F_M$, which is the intersection of $F$ with the dual weight lattice of $G$, refined by the factor $1/M$. Positive integer $M$ fixes the density of points in $F_M$. 
The symmetry of the dual weight lattice is dictated by our choice of $G$. Digital data are functions sampled on $F_M$. They may be of wildly varying densities. Therefore the flexibility in the choice of $M$ is crucial in our set up.

The results in this paper provide all the information that is needed for three types of expansions of functions, given on the fundamental region $F$ of any compact simple Lie group/Lie algebra. The fundamental region is an $n$-dimensional simplex given explicitly by its $n+1$ vertices. When the Lie group/Lie algebra is semisimple but not simple, the data needed are a straightforward concatenation of those needed for simple cases. In particular, the fundamental region is the Cartesian product of simplexes of simple constituents of the Lie group/Lie algebra.

Firstly, we consider the expansion into series of either $C-$ or $S-$functions, analog of the expansions into Fourier series in $n$ variables. Unlike the familiar case, our variables do not refer necessarily to an orthogonal basis in $\R^n$. That is just one of the cases when the underlying Lie algebra is the product $A_1\times A_1\times\cdots\times A_1$ of $n$ copies of $A_1$. Here the variables are defined relative to a basis which reflects the symmetries of the weight lattice of the Lie algebra. Hence, in all but one case, the basis is non-orthogonal.

Secondly, we consider the expansion of functions sampled on a lattice grid $F_M\subset F$ with density controlled by $M\in\N$. Such expansions are necessarily finite because the number of points $|F_M|$ is finite. The numbers $|F_M|$ are determined here for all simple Lie algebras and for all $M$.

Our third case examines the continuous interpolations of functions sampled on $F_M$. Once the corresponding finite series is found, it suffices to replace the lattice variables in the expansion functions by the continuous variables. Such a simple idea applies equally well to any dimension, to lattices of any symmetry and to grids in $F$ of any density. The interpolation is performed in the `Fourier space' rather than in the space of the data, where the best of the common interpolation techniques are usually applied. At present there is some empirical evidence that the quality of our interpolation method compares favorably with the best methods used in 2D, while being as fast as the simplest of the standard interpolations in 2D, namely the linear interpolation. In dimensions exceeding 2, there is hardly any competition to the possibilities brought forth here.

The technical data provided in this paper in full generality are as follows. First we provide the formula for the volume vol($F$) of the fundamental region of $G$, which is closely related to the order of the Weyl group $W$ of $G$.

A new result presented in this paper is the number $|F_M|$ of lattice points in the fragment $F_M$ for all $1\leq M<\infty$ and for all simple Lie groups $G$. The number $|F_M|$  is used in the discrete orthogonality relation of the orbit functions when those are sampled from the points of $F_M$.

The second result presented in this paper is the determination of the sets $\Lambda_M$ of $C-$ and $S-$functions which are pairwise orthogonal on the point sets $F_M$. Such a set consists of points of the weight lattice $P$ of $G$. In fact there is an unlimited number of such sets in $P$. Our aim is to describe the unique lowest set.

The third result presented in this paper is a method for finding a number indispensable to the definition of the inner product of functions sampled on the points of $F_M$. The number is equal to the number of elements in a conjugacy class of elements of finite order on the maximal torus of a simple Lie group. That number, and also an implicit method for determining it, have already been found in \cite{MP1}. The present prescription is explicit and easy to use in all cases.

In order to determine the numbers $|F_M|$ and the sets $\Lambda_M$, we must review the orthogonality properties of orbit functions in a somewhat wider context than in \cite{MP1,MP2}.

Reasons for which the results presented should be of interest, beyond our own motivations, are as follows. 

The volume of $F$ for each $G$ is needed when the continuous orthogonality of orbit functions is used. Its relation to the order of the Weyl group is of independent interest.

The points of $F_M$ are obtained in terms of barycentric coordinates $[s_0,s_1,\dots,s_n]$ satisfying a certain sum rule \eqref{FM}. The subset of $F_M$, subject to the additional requirement that $\gcd\{s_0,s_1,\dots,s_n\}=1$, consists of the representatives of the conjugacy classes of elements of adjoint order $M$ in the corresponding compact simple Lie group. Counting such conjugacy classes in certain Lie groups was the subject of the papers \cite{D1,D11,D2}.

The sets $\Lambda_M$ and $F_M$ are in one-to-one correspondence. Most of the properties of elements of $F_M$ are therefore reflected in the corresponding elements of $\Lambda_M$. 

The method used here to determine the number of conjugate points to every point of $F_M$ can be extracted from much more elaborate techniques exploited in \cite{MP4} for describing the faces of all dimensions of the proximity ('Voronoi') cells of the root lattices of $G$.

Pertinent standard properties of simple Lie algebras are recalled in Section~2. A general formula for the volume of $F$ is given there. In Section~3 the lattice grids $F_M$ and the set $\Lambda_M$ are described and the number of their points is found. Section~4 contains a description of the $C-$ and $S-$functions and their properties pertinent for our goals. In view of the preceding results, Section~5 is devoted to the detailed description of the discrete orthogonality of $C-$ and $S-$function. The distinction between the two is brought out. Comments and interesting follow-up questions are in the last section.

\section{Pertinent properties of Lie groups and Lie algebras}

\subsection{Definitions and notations}\

Consider the Lie algebra of the compact simple Lie group $G$ of rank $n$, with the set of simple roots $\Delta=\{\al_1,\dots,\al_n\}$, spanning the Euclidean space $\R^n$.

\begin{figure}[!ht]
\resizebox{16.5cm}{!}{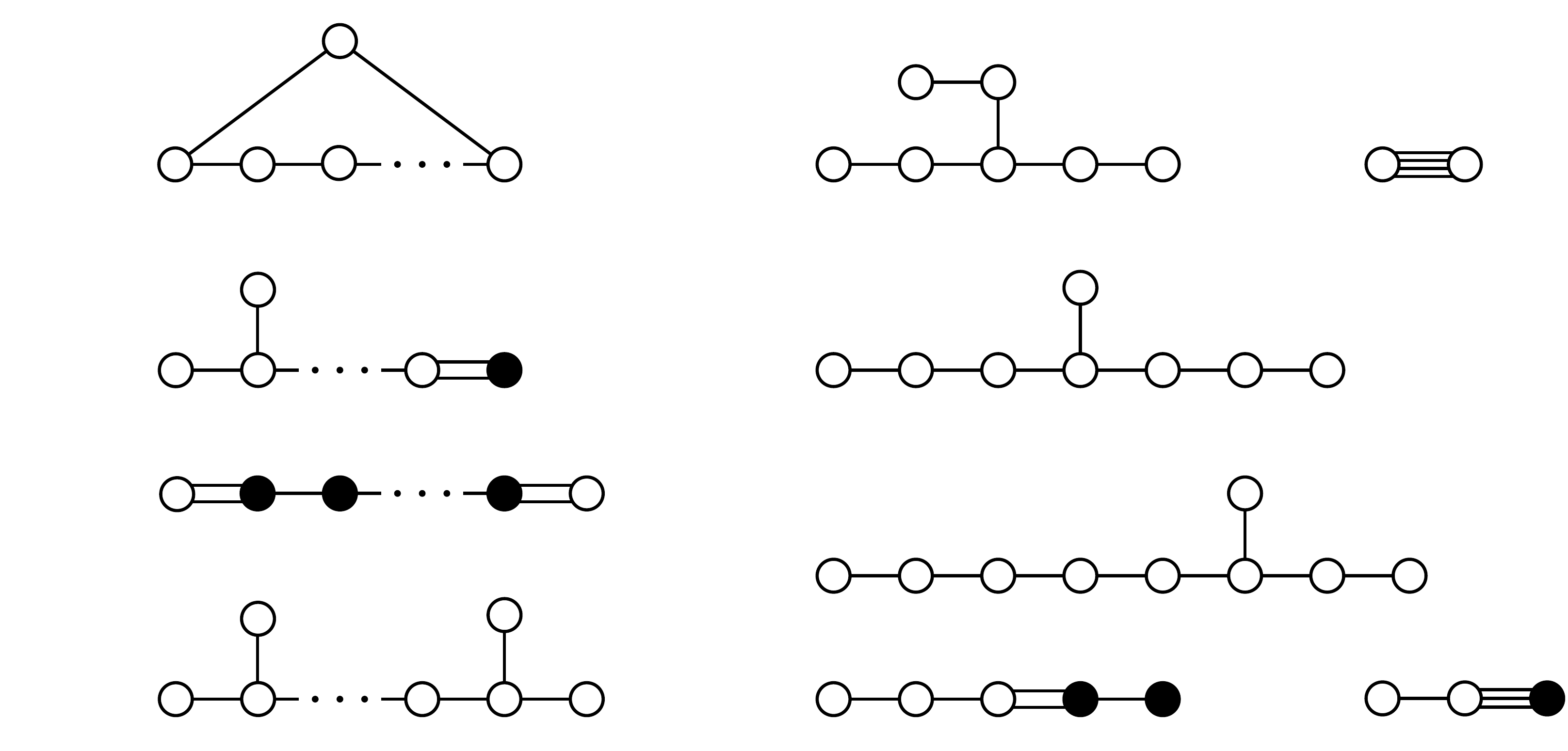}
\caption{ The extended Coxeter-Dynkin diagrams of simple Lie algebras and their numbering. The extension root carries number 0.}\label{diaggg}
\end{figure}

By uniform and standard methods for $G$ of any type and rank, one determines from $\Delta$ a number of related quantities and virtually all the properties of $G$. We make use of the following ones.
\begin{itemize}
\item
The highest root $\xi\equiv -\al_0=m_1\al_1+\dots+m_n\al_n$. Here the coefficients $m_j$ are known positive integers also called the marks of $G$.

\item
The Coxeter number $m=1+m_1+\dots+m_n$ of $G$.

\item
The Cartan matrix $C$
\begin{equation*}
 C_{ij}=\frac{2\sca{\al_i}{\al_j} }{\sca{\al_j}{\al_j}},\q i,j\in\{1,\dots,n\}.
\end{equation*}

\item
The order $c$ of the center of $G$,
\begin{equation}\label{Center}
 c=\det C.
\end{equation}
\item
The root lattice $Q$ of $G$,
 \begin{equation}\label{dPPP}
 Q=\set{\al\in \R^n}{\sca{\al}{\om^{\vee}}\in\Z,\,\forall\om^{\vee}\in P^{\vee}}=\Z\al_1+\dots+\Z\al_n\,.
\end{equation}

\item
The $\Z$-dual lattice to $Q$,
\begin{equation*}
 P^{\vee}=\set{\om^{\vee}\in \R^n}{\sca{\om^{\vee}}{\al}\in\Z,\, \forall \al \in \Delta}=\Z \om_1^{\vee}+\dots +\Z \om_n^{\vee}\,.
 \end{equation*}

\item
The dual root lattice
\begin{equation*}
 Q^{\vee}=\Z \al_1^{\vee}+\dots +\Z \al^{\vee}_n\,,\quad\text{where}\quad
     \al^{\vee}_i=\frac{2\al_i}{\sca{\al_i}{\al_i}}\,.
\end{equation*}

\item
The extended Coxeter-Dynkin diagram ($\mathrm{DD}$) of $G$ -- this diagram describes the system of the simple roots~$\Delta$ together with the highest root~$\xi=-\al_0$. The~$k$-th node corresponds to the vector~$\alpha_k,\, k=0,\dots,n$. Direct links between two nodes indicate absence of orthogonality between the pair of the corresponding vectors. Single, double and triple vertices imply that relative angles between these vectors are~$2\pi/3,\, 3\pi /4, \, 5\pi /6$, respectively. Quadruple vertex, which appears for the case~$A_1$ only, denotes the fact that the highest root and the simple root coincide. Colors of the nodes indicate relative length of~$\alpha_k$. For the cases $B_n, C_n$ and $F_4$ the squared length of the black node (short root) is half of the squared length of the white node (long root). For~$G_2$, the squared length of the black node is one third of the squared length of the white node. We also use the standard additional convention for the squared lengths of the white nodes~$\alpha$
$$
\sca{\alpha}{\alpha}=2.
$$
The original (non-extended) Coxeter-Dynkin diagram of $G$ can be recovered from extended $\mathrm{DD}$ by omitting the extension $0$-node and adjacent edges.

The extended $\mathrm{DD}$'s of all simple Lie algebras are shown in Figure~\ref{diaggg}.
\end{itemize}

\subsection{Weyl group and affine Weyl group}\

The properties of Weyl groups and affine Weyl groups can be found for example in \cite{H2,BB}.
The finite Weyl group $W$ is generated by $n$ reflections $r_\al$, $\al\in\Delta$, in $(n-1)$-dimensional `mirrors' orthogonal to simple roots.
\begin{equation*}
r_{\al_i}a\equiv r_i a=a-\frac{2\sca{a}{\al_i} }{\sca{\al_i}{\al_i}}\al_i\,,
\qquad a\in\R^n\,.
\end{equation*}

The infinite affine Weyl group $W^{\mathrm{aff}}$ is generated by translations by elements of~$Q^\vee$ and by~$W$. More precisely, $W^{\mathrm{aff}}$ is the semidirect product of the Abelian group of translations $Q^\vee$ and of the Weyl group~$W$.
\begin{equation}\label{direct}
 W^{\mathrm{aff}}= Q^\vee \rtimes W.
\end{equation}
Equivalently, $W^{\mathrm{aff}}$ is generated by reflections $r_i$ and reflection $r_0$, where
\begin{equation*}
r_0 a=r_\xi a + \frac{2\xi}{\sca{\xi}{\xi}}\,,\qquad
r_{\xi}a=a-\frac{2\sca{a}{\xi} }{\sca{\xi}{\xi}}\xi\,,\qquad a\in\R^n\,.
\end{equation*}

The fundamental region $F$ of $W^{\mathrm{aff}}$ is the convex hull of the points $\left\{ 0, \frac{\om^{\vee}_1}{m_1},\dots,\frac{\om^{\vee}_n}{m_n} \right\}$:
\begin{align}
  F &=\setb{y_1\om^{\vee}_1+\dots+y_n\om^{\vee}_n}{y_0,\dots, y_n\in \R_0^+, \, y_0+y_1 m_1+\dots+y_n m_n=1  } \label{deffun} \\
&= \setb{a\in \R^n}{\sca{a}{\al}\geq 0,\forall \al\in\Delta,\sca{a}{\xi}\leq 1  }\nonumber
\end{align}
and
\begin{equation*}
  W^{\mathrm{aff}} F =\R^n .
\end{equation*}

Since $F$ is a fundamental region of $W^{\mathrm{aff}}$, we have:
\begin{enumerate}
\item For any $a\in \R^n$ there exists $a'\in F$, $w\in W$ and $q^{\vee}\in Q^{\vee}$ such that
\begin{equation}\label{fun1}
 a=wa'+q^{\vee}.
\end{equation}

\item If $a,a'\in F$ and $a'=w^{\mathrm{aff}}a$, $w^{\mathrm{aff}}\in W^{\mathrm{aff}}$ then $a=a'$, i.e. if there exist $w\in W$ and $q^{\vee}\in Q^{\vee}$ such that $a'=wa+q^{\vee}$ then
\begin{equation}\label{fun2}
 a'=a=wx+q^{\vee}.
\end{equation}

\item
Consider a point
$a=y_1\om^{\vee}_1+\dots+y_n\om^{\vee}_n\in F$,  such that
$y_0+y_1 m_1+\dots+y_n m_n=1$.
 The isotropy group
 \begin{equation}\label{stab}
 \mathrm{Stab}_{W^{\mathrm{aff}}}(a) = \setb{w^{\mathrm{aff}}\in W^{\mathrm{aff}}}{w^{\mathrm{aff}}a=a}
\end{equation}
of the point $a$ is trivial, $\mathrm{Stab}_{W^{\mathrm{aff}}}(a)=1$, if $a\in \mathrm{int}(F)$, i.e. all $y_i>0$, $i=0,\dots,n$. Otherwise the group $\mathrm{Stab}_{W^{\mathrm{aff}}}(a)$
is generated by such $r_i$ for which $y_i=0$, $i=0,\dots,n$. The Coxeter--Dynkin diagram of $\mathrm{Stab}_{W^{\mathrm{aff}}}(a)$ is obtained as a subgraph of the extended $\mathrm{DD}$ by omitting such vertices $\al_i$ (and adjacent edges) for which $y_i>0$, $i=0,\dots,n$. Since there always exists some $y_i>0$, $i=0,\dots,n$ this subgraph is always proper.
\end{enumerate}

\subsection{Volume of $F$}\

The volumes of fundamental domains $\mathrm{vol}(F)$ often appear in an implicit form in the literature and are commonly used for the calculation of the order of the Weyl group~\cite{Bour,H2}. For applications, it may be also useful to calculate explicit values of these volumes which we summarize in Table~\ref{tab1}. Taking into account the standard convention for the lengths of the root vectors, one can use for the calculation of $\mathrm{vol}(F)$ the following formula.
\begin{tvr}
The volume $\mathrm{vol}(F)$ of the fundamental domain $F\subset \R^n$ is equal to
\begin{equation*}
 \mathrm{vol}(F)=\frac{1}{n!}\frac{1}{m_1\cdots m_n}\sqrt{\frac{2}{\langle \alpha_1,\alpha_1\rangle}\cdots\frac{2}{\langle \alpha_n,\alpha_n\rangle}}\,  c^{-\frac{1}{2}}.
\end{equation*}
\end{tvr}
\begin{proof}
Suppose we have standard orthonormal basis $\chi$ of $\R^n$. The map which transforms standard $n-$simplex into $F$ is determined by the matrix $\omega^{\vee}_m$ given by the coordinates of the vectors $\omega^{\vee} _i/ m_i,i\in\{1,\dots,n\}$ in the basis $\chi$. Since the volume of the standard $n$-simplex is $1/n!$, we have
\begin{equation*}
 \mathrm{vol}(F)=\frac{1}{n!}| \det \omega^{\vee}_m |.
\end{equation*}
Let $\omega^{\vee}, \al^{\vee}$ denote the matrices given by the coordinates (in columns) of the vectors  $\omega^{\vee} _i$ and $\al^{\vee} _i$, respectively and let $G(\al^{\vee}_1,\dots\al^{\vee}_n)$ denote the Gram matrix of the basis $\al^{\vee}_1,\dots\al^{\vee}_n$, i. e. $G(\al^{\vee}_1,\dots\al^{\vee}_n)_{ij}=\sca{\al^{\vee}_i}{\al^{\vee}_j}$. Then $(\al^{\vee})^{T}\al^{\vee}=G(\al^{\vee}_1,\dots\al^{\vee}_n)$ and $c\det \omega^{\vee}= \det\al^{\vee} $. Clearly, the relation $$\det G(\al^{\vee}_1,\dots\al^{\vee}_n)=\frac{2}{\langle \alpha_1,\alpha_1\rangle}\cdots\frac{2}{\langle \alpha_n,\alpha_n\rangle}c  $$ also holds. We now have

$$| \det \omega^{\vee}_m |=\frac{1}{m_1\cdots m_n}| \det \omega^{\vee}|= \frac{1}{m_1\cdots m_n}c^{-1}| \det \al^{\vee}|=\frac{1}{m_1\cdots m_n}c^{-1}\sqrt{\det G(\al^{\vee}_1,\dots\al^{\vee}_n)}.$$
\end{proof}

\subsection{Action of $W$ on the maximal torus $\R^n/Q^{\vee}$}\

If we have two elements $a,a'\in \R^n$ such that $a'-a=q^{\vee}$, with $q^{\vee}\in Q^{\vee}$ then for $w\in W$ we have $wa-wa'=wq^{\vee}\in Q^{\vee}$, i.e. we have a natural action of $W$ on the torus $\R^n/Q^{\vee}$. For $x\in \R^n/Q^{\vee}$ we denote the isotropy group
 \begin{equation*}
\mathrm{Stab} (x)=\set{w\in W}{wx=x}
\end{equation*}
and its order by $h_x\equiv |\mathrm{Stab} (x)|$. We denote the orbit
 \begin{equation*}
W x=\set{wx\in \R^n/Q^{\vee} }{w\in W}
\end{equation*}
and its order by $\ep(x)\equiv |Wx|$. Clearly we have
\begin{equation}\label{ep}
\ep(x)=\frac{|W|}{h_x}.
\end{equation}

\begin{tvr}\begin{enumerate}
\item For any $x\in \R^n/Q^{\vee}$ there exists $x'\in F \cap \R^n/Q^{\vee} $ and $w\in W$ such that
\begin{equation}\label{rfun1}
 x=wx'.
\end{equation}

\item If $x,x'\in F \cap \R^n/Q^{\vee} $ and $x'=wx$, $w\in W$ then
\begin{equation}\label{rfun2}
 x'=x=wx.
\end{equation}
\item If $x\in F \cap \R^n/Q^{\vee} $, i.e. $x=a+Q^{\vee}$, $a\in F$ then
\begin{equation}\label{rfunstab}
\mathrm{Stab} (x) \cong \mathrm{Stab}_{W^{\mathrm{aff}}}(a).
\end{equation}
\end{enumerate}
\end{tvr}
\begin{proof}
\begin{enumerate}
 \item Follows directly from (\ref{fun1}).
\item Follows directly from (\ref{fun2}).
\item We have from (\ref{direct}) that for any $w^{\mathrm{aff}}\in W^{\mathrm{aff}}$ there exist unique $w\in W$ and unique shift $T(q^{\vee})$ such that $w^{\mathrm{aff}}=T(q^{\vee})w$. Define map $\psi: \mathrm{Stab}_{W^{\mathrm{aff}}}(a)\map W $ for $w^{\mathrm{aff}}\in \mathrm{Stab}_{W^{\mathrm{aff}}}(a)$ by
\begin{equation*}
\psi(w^{\mathrm{aff}})=\psi(T(q^{\vee})w)=w.
\end{equation*}
Since the relation
\begin{equation*}
\psi(T(q_1^{\vee})w_1T(q_2^{\vee})w_2)=\psi(T(q_1^{\vee}+w_1q_2^{\vee})w_1w_2)=w_1w_2
\end{equation*}
holds, $\psi$ is indeed a homomorphism. If $a=w^{\mathrm{aff}}a=wa+q^{\vee}$ then $a-wa=q^{\vee}\in Q^{\vee}$, i.e. we obtain $w\in \mathrm{Stab} (x)$ and vice versa. Thus, $\psi (\mathrm{Stab}_{W^{\mathrm{aff}}}(a))=\mathrm{Stab}(x)$ holds. We also have
\begin{equation*}
\mathrm{ker}\,\psi= \left\{ T(q^{\vee})\in \mathrm{Stab}_{W^{\mathrm{aff}}}(a) \right\}=1.
\end{equation*}
\end{enumerate}

\end{proof}

\subsection{Dual Lie algebra}\

The set of simple dual roots $\Delta^{\vee}=\{\al^{\vee}_1,\dots , \al^{\vee}_n\}$ is a system of simple roots of some simple Lie algebra. The system
$\Delta^{\vee}$ also spans Euclidean space $\R^n$.

The dual system $\Delta^{\vee}$ determines:
\begin{itemize}
\item
The highest dual root $\eta\equiv -\al_0^{\vee}= m_1^{\vee}\al_1^{\vee} + \dots + m_n^{\vee} \al_n^{\vee}$. Here the coefficients $m^{\vee}_j$ are called the dual marks of $G$. The marks and the dual marks are summarized in Table \ref{tabmarks}.
{
\begin{table}
\begin{tabular}{|c||c||c|}
\hline
Type & Marks $m_1,\dots,m_n$ & Dual marks $m_1^\vee,\dots,m_n^\vee$ \\
\hline\hline
$A_n\ (n\geq1)$ & $1,1,\dots,1$ & $1,1,\dots,1$ \\ \hline
$B_n\ (n\geq3)$ & $1,2,2,\dots,2$ & $2,2,\dots,2,1$ \\ \hline
$C_n\ (n\geq2)$ & $2,2,\dots,2,1$ & $1,2,2,\dots,2$  \\ \hline
$D_n\ (n\geq4)$ & $1,2,\dots,2,1,1$ & $1,2,\dots,2,1,1$ \\ \hline
$E_6$ & $1,2,3,2,1,2$ &  $1,2,3,2,1,2$ \\ \hline
$E_7$ & $2,3,4,3,2,1,2$ & $2,3,4,3,2,1,2$ \\ \hline
$E_8$ & $2,3,4,5,6,4,2,3$ & $2,3,4,5,6,4,2,3$ \\ \hline
$G_2$ & $2,3$ & $3,2$ \\ \hline
$F_4$ & $2,3,4,2$ & $2,4,3,2$ \\ \hline
\end{tabular}
\medskip

\caption{ The marks and the dual marks of simple Lie algebras. Numbering of simple roots as on Fig.~\ref{diaggg} is used.}\label{tabmarks}
\end{table}
}

\item
The dual Cartan matrix $C^{\vee}$
 \begin{equation*}
 C^{\vee}_{ij}=\frac{2\sca{\al^{\vee}_i}{\al^{\vee}_j} }{\sca{\al^{\vee}_j}{\al^{\vee}_j}}=C_{ji},\q i,j\in\{1,\dots,n\}.
\end{equation*}

\item
The dual root lattice
 \begin{equation*}
 Q^{\vee}=\Z \al^{\vee}_1+\dots +\Z \al^{\vee}_n.
\end{equation*}

\item
The root lattice
\begin{equation*}
 Q=\Z \al_1+\dots +\Z \al_n\,,\quad\text{where}\quad \al_i=\frac{2\al_i^{\vee}}{\sca{\al_i^{\vee}}{\al_i^{\vee}}}.
\end{equation*}

\item The $\Z$-dual lattice
\begin{equation*}
 P=\set{\om\in \R^n}{\sca{\om}{\al^{\vee}}\in\Z,\, \forall \al^{\vee} \in \Delta^{\vee}}=\Z \om_1+\dots +\Z \om_n.
 \end{equation*}
\item
The extended dual Coxeter-Dynkin diagram $(\mathrm{DD}^{\vee})$ of $G$ describe the system of dual roots $\Delta^{\vee}$ together with the highest dual root $\eta=-\al^{\vee}_0$. The $k$-th node corresponds to the vector $\alpha^{\vee}_k,\, k=0,\dots,n$. The rules related to relative angles and lengths of vectors are the same as for extended diagrams. Note that the squared lengths of the white nodes (long dual roots) $\alpha^{\vee}$ are determined by the formula
$$
\sca{\al^{\vee}}{\al^{\vee}}=\frac{4}{\sca{\alpha}{\al}},
$$
and in general $\sca{\al^{\vee}}{\al^{\vee}}\neq 2$. For the extended $\mathrm{DD}^{\vee}$ the following cases may occur:
\begin{enumerate}
\item For the algebras $A_n$, $D_n$, $E_6$, $E_7$ and $E_8$ extended $\mathrm{DD}^{\vee}$ coincide with the corresponding extended $\mathrm{DD}$ in Fig. \ref{diaggg}.
\item Extended $\mathrm{DD}^{\vee}$ of $B_n$, $n\geq 3$ is the extended $\mathrm{DD}$ of $C_n$ and vice versa.
\item Extended $\mathrm{DD}^{\vee}$ of $C_2$ with unchanged numbering of the nodes
\begin{center}
\vspace{6pt}
\resizebox{2.0cm}{!}{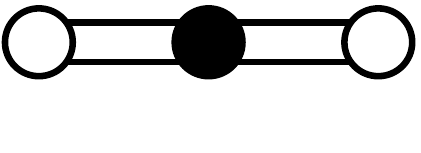}
\end{center}
\item Extended $\mathrm{DD}^{\vee}$ of $G_2$ with unchanged numbering of the nodes
\begin{center}
\vspace{6pt}
\resizebox{2.0cm}{!}{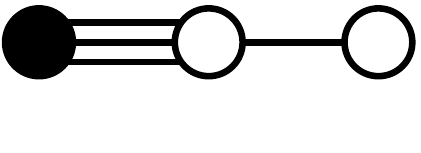}
\end{center}
\item Extended $\mathrm{DD}^{\vee}$ of $F_4$ with unchanged numbering of the nodes
\begin{center}
\vspace{6pt}
\resizebox{3.7cm}{!}{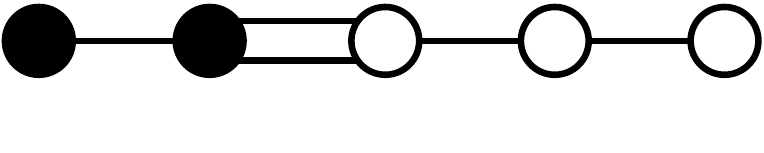}
\end{center}
\end{enumerate}
\end{itemize}
\subsection{Dual affine Weyl group}\

The group generated by reflections with respect to dual roots $r_{\al^{\vee}}=r_\al$, $\al^{\vee}\in\Delta^{\vee}$ coincides with the
Weyl group $W$. Dual affine Weyl group $\widehat{W}^{\mathrm{aff}}$ is generated by shifts from $Q$ and Weyl group $W$.
Moreover, $\widehat{W}^{\mathrm{aff}}$ is a semidirect product of group of shifts  $Q$ and Weyl group $W$
\begin{equation}\label{directd}
 \widehat{W}^{\mathrm{aff}}= Q \rtimes W.
\end{equation}
Equivalently, $\widehat{W}^{\mathrm{aff}}$ is generated by reflections $r_i$ and reflection $r_0^{\vee}$, where
\begin{equation*}
r_0^{\vee} a=r_{\eta} a + \frac{2\eta}{\sca{\eta}{\eta}}, \q r_{\eta}a=a-\frac{2\sca{a}{\eta} }{\sca{\eta}{\eta}}\eta,\q a\in\R^n.
\end{equation*}

The fundamental region $F^\vee$ of $\widehat{W}^{\mathrm{aff}}$ is the convex hull
 of the vertices $\left\{ 0, \frac{\om_1}{m^{\vee}_1},\dots,\frac{\om_n}{m^{\vee}_n} \right\}$:

\begin{align}
  F^\vee &=\setb{z_1\om_1+\dots+z_n\om_n}{z_0,\dots, z_n\in \R_0^+, \, z_0+z_1 m_1^{\vee}+\dots+z_n m^{\vee}_n=1  }\label{defdfun} \\
&= \setb{a\in \R^n}{\sca{a}{\al^{\vee}}\geq 0,\forall \al^{\vee}\in\Delta^{\vee},\sca{a}{\eta}\leq 1  }\nonumber
\end{align}
and
\begin{equation*}
  \widehat{W}^{\mathrm{aff}} F^\vee =\R^n.
\end{equation*}

Since $F^\vee$ is a fundamental region, we have:
\begin{enumerate}
\item For any $a\in \R^n$ there exists $a'\in F^\vee$, $w\in W$ and $q\in Q$ such that
\begin{equation}\label{ddfun1}
 a=wa'+q.
\end{equation}

\item If $a,a'\in F^\vee$ and $a'=w^{\mathrm{aff}}a$, $w^{\mathrm{aff}}\in \widehat{W}^{\mathrm{aff}}$ then $a=a'$, i.e. if there exist $w\in W$ and $q\in Q$ such that $a'=wa+q$ then
\begin{equation}\label{dfun2}
 a'=a=wa+q.
\end{equation}
\item Consider the point $a=z_1\om_1+\dots+z_n\om_n\in F^\vee$ such that $z_0+z_1 m_1^{\vee}+\dots+z_n m^{\vee}_n=1$. The isotropy group
\begin{equation}\label{stabdual}
 \mathrm{Stab}_{\widehat{W}^{\mathrm{aff}}}(a) = \setm{w^{\mathrm{aff}}\in \widehat{W}^{\mathrm{aff}}}{w^{\mathrm{aff}}a=a}
\end{equation}
of the point $a$ is trivial, $ \mathrm{Stab}_{\widehat{W}^{\mathrm{aff}}}(a)=1$, if $a\in \mathrm{int}(F^\vee)$, i.e. all $z_i>0$, $i=0,\dots,n$. Otherwise the group $\mathrm{Stab}_{\widehat{W}^{\mathrm{aff}}}(a)$
is generated by such $r^{\vee}_i$ for which $z_i=0$, $i=0,\dots,n$. The Coxeter--Dynkin diagram of $\mathrm{Stab}_{\widehat{W}^{\mathrm{aff}}}(a)$ is obtained as a subgraph of extended $\mathrm{DD}^{\vee}$ by omitting such vertices $\al^{\vee}_i$ (and adjacent edges) for which $z_i>0$, $i=0,\dots,n$. Since there always exists some $z_i>0$, $i=0,\dots,n$ this subgraph is always proper.
\end{enumerate}

\section{Grids $F_M$ and $\Lambda_M$ }

\subsection{Grid $F_M$}\

The grid $F_M$ is the finite fragment of the lattice $\frac{1}{M}P^{\vee}$ which is found inside of $F$. Suppose we have fixed $M\in \N$ and consider $W$-invariant group $\frac{1}{M}P^{\vee}/Q^{\vee}$
of the finite order
\begin{equation}\label{grupa}
\abs{\frac{1}{M}P^{\vee}/Q^{\vee}}=cM^n.
\end{equation}

We define the grid $F_M$ as such cosets from $\frac{1}{M}P^{\vee}/Q^{\vee}$ which have representative element in the fundamental domain $F$:
\begin{equation*}
F_M\equiv\frac{1}{M}P^{\vee}/Q^{\vee}\cap F.
 \end{equation*}

From the relation (\ref{rfun1}) we have that
\begin{equation}\label{WFM}
 WF_M = \frac{1}{M}P^{\vee}/Q^{\vee}.
 \end{equation}

We have immediately from (\ref{deffun}) that the set $F_M$, or more precisely its representative points, can be identified as
\begin{equation}\label{FM}
 F_M = \setb{\frac{s_1}{M}\om^{\vee}_1+\dots+\frac{s_n}{M}\om^{\vee}_n}{s_0,s_1,\dots ,s_n \in \Z^{\geq 0},\, s_0+s_1m_1+\dots + s_n m_n=M}.
 \end{equation}

Another interpretation of the points of $F_M$ is as points representing the conjugacy classes of finite abelian subgroup of the maximal torus which is generated by the elements of order $M$ of the torus. The order of that group can be determined after the number of elements of $F_M$ is found in the next subsection.
\subsection{Number of elements of $F_M$}\

The number of elements of $F_M$ could be obtained by putting together the results of Djokovi\'c \cite{D1,D11}, namely the numbers corresponding of all divisors of $M$. We proceed in another way by determining the number directly.
\begin{tvr}\label{AN}
Let $K,l\in \Z^{\geq 0}$. Then the number of solutions of the equation
\begin{equation}\label{equation}
 a_0+a_1+\dots + a_l= K,\q a_0,\dots,a_l\in \Z^{\geq 0}
\end{equation}
is equal to $\comb{l+K}{l}$.
\end{tvr}
\begin{proof}
We proceed by induction on $l$. If $l=0$ then the equation $a_0=K$ has exactly one solution. We arrange the solutions of (\ref{equation}) into $K+1$ disjoint subsets according the value of $a_l=0,\dots,K$. According to the assumption, the number of solutions in each of these subsets is equal to $\comb{l-1+K-a_l}{l-1}$. Then we have
\begin{equation*}
 \sum_{a_l=0}^K \comb{l-1+K-a_l}{l-1} = \sum_{k=0}^K \comb{l-1+k}{l-1}=\comb{l+K}{l}.
\end{equation*}
\end{proof}

\begin{tvr}\label{AN2}
Let $n,m_1,\dots,m_n,M\in \N$, denote $L=\mathrm{lcm} (m_1,\dots,m_n)$ and suppose
\begin{align*}
M&=Lk+l,\q k,l\in \Z^{\geq 0},\, l<L \\
(n+1)L-( 1+m_1+\dots +m_n)&=LN+N' \q N,N'\in \Z^{\geq 0},\, N'<L.
\end{align*}
Then the number of solutions of the equation
\begin{equation}\label{equation2}
 a_0+m_1a_1+\dots + m_n a_n= M,\q a_0,\dots,a_n\in \Z^{\geq 0}
\end{equation}
is equal to $\sum_{i=0}^{N}d_{li}\comb{n-i+k}{n},$
where $d_{li}$ is the number of solutions of the equation
\begin{equation}\label{equation3}
 l_0+m_1l_1+\dots + m_n l_n= Li+l,\q l_0\in\{0,\dots, L-1\},\,l_1\in\{0,\dots, \frac{L}{m_1}-1\},\dots,l_n\in\{0,\dots, \frac{L}{m_n}-1\}.
\end{equation}
\end{tvr}
\begin{proof}
Suppose we have fixed $M=Lk+l,$ $k,l\in \Z^{\geq 0},\, l<L$ and denote by $[a_0,a_1,\dots,a_n]$ a solution of (\ref{equation2}).
Consider the mapping $$\,[a_0,a_1,\dots,a_n]\mapsto[ [l_0,l_1,\dots,l_n],[k_0,k_1,\dots,k_n] ]$$ via the relations
\begin{align}\label{decomp}
a_0&=Lk_0+l_0,\q k_0,l_0\in \Z^{\geq 0},\, l_0<L \nonumber \\
a_1&=\frac{L}{m_1}k_1+l_1,\q k_1,l_1\in \Z^{\geq 0},\, l_1<\frac{L}{m_1}\nonumber \\
& \vdots \\
a_n&=\frac{L}{m_n}k_n+l_n,\q k_n,l_n\in \Z^{\geq 0},\, l_n<\frac{L}{m_n}. \nonumber
\end{align}
Substituing (\ref{decomp}) into (\ref{equation2}) we obtain that $[l_0,l_1,\dots,l_n]$ is a solution of (\ref{equation3}) for some $i\in \{0,1,\dots,N\}$ such that
\begin{equation}\label{equation4}
 k_0+k_1+\dots + k_n = k-i.
\end{equation}
Conversely, for fixed $i\in \{0,1,\dots,N\}$, each solution $[k_0,k_1,\dots,k_n]$ of (\ref{equation4}) together with a solution of (\ref{equation3}) leads to some solution of (\ref{equation2}), i.e. we have a decomposition of the solutions of (\ref{equation2}) into disjoint subsets $S_i,\,i\in \{0,1,\dots,N\}$. Taking into account that the number of solutions of (\ref{equation4}) is $\comb{n+k-i}{n}$, we have $ |S_i|=d_{li}\comb{n+k-i}{n}$.
\end{proof}
\begin{thm}\label{numAn}
The numbers of points of the grid $F_M$ of Lie algebras $A_n$, $B_n$, $C_n$, $D_n$ are given by the following relations.
\begin{enumerate}
 \item $A_n,\,n\geq 1$ $$|F_M(A_n)|=\begin{pmatrix}n+M\\ n \end{pmatrix}$$
\item  $C_n,\,n\geq 2$
\begin{eqnarray*}
|F_{2k}(C_n)|=& \begin{pmatrix}n+k \\ n \end{pmatrix}+\begin{pmatrix}n+k-1 \\ n \end{pmatrix}\\
|F_{2k+1}(C_n)|=& 2\begin{pmatrix}n+k \\ n \end{pmatrix} \end{eqnarray*}
\item $B_n,\,n\geq 3$ $$|F_M(B_n)|=|F_M(C_n)|$$
\item $D_n,\,n\geq 4$\begin{eqnarray*}
|F_{2k}(D_n)|=& \begin{pmatrix}n+k \\ n \end{pmatrix}+6\begin{pmatrix}n+k-1 \\ n \end{pmatrix}+\begin{pmatrix}n+k-2 \\ n \end{pmatrix}\\
|F_{2k+1}(D_n)|=& 4\begin{pmatrix}n+k \\ n \end{pmatrix}+4\begin{pmatrix}n+k-1 \\ n \end{pmatrix} \end{eqnarray*}
\end{enumerate}
 \end{thm}
\begin{proof}
\begin{enumerate}
 \item Follows directly from Proposition \ref{AN}.
\item We have $m_1=2,m_2=2,\dots,m_n=1$ and thus, $L=2$ and Coxeter number $m=2n$, $N=1$.
\begin{enumerate}[(a)]
 \item $M=2k$. According to Proposition \ref{AN2} we have to analyze the number of solutions of the equation
\begin{equation*}
 l_0+ l_n= 2i,\q l_0\in\{0,1\},l_n\in\{0,1\}.
\end{equation*}
for $i\in\{0,1\}$. This equation has one solution $[0,0]$ for $i=0$ and one solution $[1,1]$ for $i=1$.
\item $M=2k+1$. The equation
\begin{equation*}
 l_0+ l_n= 2i+1,\q l_0\in\{0,1\},l_n\in\{0,1\}.
\end{equation*}
 has two solutions $[0,1],\,[1,0]$ for $i=0$ and no solution for $i=1$.
\end{enumerate}

\end{enumerate}
The proof of the cases $B_n,\,D_n$ is analogous.
\end{proof}

There is a convenient way of writing the expressions for $|F_M|$. For that we introduce, for a given Lie algebra, a matrix $$R=(d_{li})\q l\in\{0,\dots,L-1\}, i\in\{0,\dots,N\},$$ where the numbers $d_{li}$ and the integers $L$, $N$ were defined in Proposition~\ref{AN2}. The values of $L,N$ are summarized in Table~\ref{tab1}. The matrix $R$ then determines the number of points of $F_M$ via the relation
\begin{equation}\label{numpoin}
 |F_{Lk+l}|=\sum_{i=0}^{N}R_{li}\comb{n-i+k}{n}.
\end{equation}
The matrices $R$ for all simple Lie algebras are listed in Appendix.
\begin{example}\label{E8num}
We demonstrate the application of formula (\ref{numpoin}) for the case $E_8$. We take the sequence of numbers in the second row of the matrix $R(E_8)$ in Appendix, i.e. the numbers $1,1520913,151753498,\dots $ and calculate
\begin{align*}
|F_{60k+1}(E_8)|=& \comb{8+k}{8}+1520913\comb{7+k}{8}+151753498\comb{6+k}{8}+1668997388\comb{5+k}{8} \\
 +& 4262344515\comb{4+k}{8}+3044716089\comb{3+k}{8}+571669092\comb{2+k}{8}\\ +&18981162\comb{1+k}{8}
+17342\comb{k}{8}=\\ =&{\frac {1}{84}} \left( 2k+1 \right)\left( 3k+1 \right) \left( 5k+1 \right)\left( 5k+2 \right) \left( 10k+3 \right)\left( 15k+2 \right) \left( 30k+1 \right) \left( 30k+7 \right)
\end{align*}
Thus we have
\begin{align*}
|F_{1}(E_8)|=&1 \\
|F_{61}(E_8)|=&1520922\\
|F_{121}(E_8)|=&165441760\\
|F_{181}(E_8)|=&3103220120\\
&\vdots
\end{align*}
The polynomial $|F_{60k+1}(E_8)|$ has real roots, and it could written in factorized form. The following polynomial can be calculated similarly but its roots are not real, therefore we do not factorize it.
\begin{align*}
|F_{60k+30}(E_8)|&=\frac{1}{56}(13500000{k}^{8}+81000000{k}^{7}+211575000{k}^{6}+314212500{k}^{5}+\\
&+290172925{k}^{4}
+170628150{k}^{3}+62388155{k}^{2}+12968910k+
 1173536)
\end{align*}
and we obtain a sequence
\begin{align*}
|F_{30}(E_8)|=&20956 \\
|F_{90}(E_8)|=&20671771\\
|F_{150}(E_8)|=&780429571\\
|F_{210}(E_8)|=&9375443806\\
&\vdots
\end{align*}
\end{example}
{\footnotesize
\begin{table}
\begin{tabular}{|c|c|c|c|c|c|c|c|c|c|}
\hline
&$A_n\ (n\geq1)$&$B_n\ (n\geq3)$&$C_n\ (n\geq2)$&$D_n\ (n\geq4)$&$E_6$&$E_7$&$E_8$&$F_4$&$G_2$\\
\hline
$|W|$&$(n+1)!$&$2^nn!$&$2^nn!$&$2^{n-1}n!$&$2^73^45$&$2^{10}3^45\cdot 7$&
                                           $2^{14}3^55^27$&$2^73^2$&$12$\\
\hline
$\mathrm{vol}(F)$&$\frac{1}{n!\sqrt{n+1}}$&$\frac{1}{n!2^{n-1}}$&$\frac{1}{n!2^{\frac{n}{2}}}$&$\frac{1}{n!2^{n-2}}$&$\frac{\sqrt{3}}{2^73^45}$&$\frac{\sqrt{2}}{2^{10}3^45\cdot7}$&
                                           $\frac{1}{2^{14}3^55^27}$&$\frac{1}{2^63^2}$&$ \frac{\sqrt{3}}{12}$\\
\hline
$c$&$n+1$&$2$&$2$&$4$&$3$&$2$&
                                           $1$&$1$&$1$\\

\hline
$m$&$n+1$&$2n$&$2n$&$2n-2$&$12$&$18$&
                                           $30$&$12$&$6$\\
\hline
$L$&$1$&$2$&$2$&$2$&$6$&$12$&
                                           $60$&$12$&$6$\\
\hline
$N$&$0$&$1$&$1$&$2$&$5$&$6$&
                                           $8$&$4$&$2$\\

 \hline

\end{tabular}
\medskip

\caption{Orders of the finite Weyl groups $\abs{W}$, volumes of $F$, orders of the center of the compact Lie group, Coxeter numbers $m$, numbers $L,N$.}\label{tab1}
\end{table}
}

\subsection{Grid $F_M$ and elements of finite order in $G$}\

The grid  $F_M$ consists of points representing the elements of Ad--order $M$ and of Ad--order equal to its divisors \cite{MP3,MP1,Kac}. The elements of Ad--order $M$ only are found by additional requirement $\gcd(s_0,s_1,\dots,s_n)=1$ in~(\ref{FM}).  Each point then represents particular conjugacy class of elements of Ad--order $M$. The relation between (full) order $k$ of elements in a conjugacy class of Ad--order $M$ is given $k=Mj$, where explicit formulas for $j$ are listed in Table~6 in \cite{MP3}.

In \cite{D1,D11} were calculated explicit formulas for the number $\nu(k,G)$ of conjugacy classes of elements whose order divide $k$, in any complex semisimple Lie group $G$. We have for example in the case of $G=C_n,\,n\geq 2$
$$\nu(2k,C_n)=\nu(2k+1,C_n)=\comb{n+k}{n}.$$
\begin{example} Consider the case $C_2$. We have from \cite{MP3} that $j=2/\gcd(2,s_2)$. In Table \ref{efo} are listed $2,4,6$ and $16$ elements of $F_1(C_2),F_2(C_2),F_3(C_2)$ and $F_6(C_2)$, respectively. 
Among these $28$  elements there are $\nu(6,C_2)=\comb{2+3}{2}=10$ elements whose order $k$ divides $6$.
\end{example}
Comparing the explicit formulas for $\nu(k,G)$ from \cite{D11} with our formulas for $\abs{F_M(G)}$, one finds a similarity which suggests some relation between them. In the following proposition, we single out such cases of $G$ for which these formulas coincide.
\begin{tvr} If $ G=G_2,F_4$ or $E_8$ then for all $M\in \N$
\begin{equation*}
\abs{F_M(G)}=\nu(M,G).
\end{equation*}
\end{tvr}
\begin{proof}
For $K|M$ define the set $F_{M,K}\subset F_M$ by
\begin{equation*}\label{FMK}
 F_{M,K}= \setb{\frac{s_1}{M}\om^{\vee}_1+\dots+\frac{s_n}{M}\om^{\vee}_n}{s_0,s_1,\dots ,s_n \in \Z^{\geq 0},\, s_0+\sum_{i=1}^n m_is_i=M,\,\gcd(s_0,s_1,\dots,s_n)=K}.
 \end{equation*}
Since one can write the set $F_M$ as a disjoint union of the sets $F_{M,K}$ with $K|M$, we have
\begin{equation}\label{disFMK}
\abs{F_M}=\sum_{K|M}\abs{F_{M,K}}.
\end{equation}
There is one-to-one correspondence between the set $F_{M,K}$ and the set $F_{\frac{M}{K},1}$ via the mapping
$$F_{\frac{M}{K},1}\ni y \mapsto K y\in F_{M,K} $$ and thus we have
\begin{equation}\label{corr}
 \abs{F_{M,K}}=\abs{F_{\frac{M}{K},1}}.
\end{equation}
Since for $G=G_2,F_4$ and $E_8$ we have from \cite{MP3} that $j=1$, the number of conjugacy classes of order $k$ is equal to $\abs{F_{k,1}}$.
Thus, using (\ref{disFMK}), (\ref{corr}) we obtain
$$\abs{F_M(G)}=\sum_{K|M}\abs{F_{M,K}}=\sum_{K|M}\abs{F_{\frac{M}{K},1}}=\sum_{K|M}\abs{F_{K,1}}=\nu(M,G).$$
 \end{proof}

{\footnotesize
\begin{table}
\begin{tabular}{cccc}
     &  $[s_0,s_1,s_2]$ & $\gcd(s_0,s_1,s_2)$  & $k$ \\
\hline
$M=1$& $[1,0,0]$ & $1$ & $1$\\
     & $[0,0,1]$ & $1$ & $2$ \\
$M=2$& $[0,1,0]$ & $1$ & $2$\\
     & $[1,0,1]$ & $1$ & $4$\\
     & $[2,0,0]$ & $2$ & $-$\\
     & $[0,0,2]$ & $2$ & $-$\\
$M=3$& $[1,1,0]$ & $1$ & $3$\\
& $[1,0,2]$ & $1$ & $3$\\
& $[2,0,1]$ & $1$ & $6$\\
& $[0,1,1]$ & $1$ & $6$\\
& $[3,0,0]$ & $3$ & $-$\\
& $[0,0,3]$ & $3$ & $-$\\
$M=6$& $[4,1,0]$ & $1$ & $6$\\
& $[2,1,2]$ & $1$ & $6$\\
& $[0,1,4]$ & $1$ & $6$\\
& $[5,0,1]$ & $1$ & $12$\\
& $[3,1,1]$ & $1$ & $12$\\
& $[1,2,1]$ & $1$ & $12$\\
& $[1,1,3]$ & $1$ & $12$\\
& $[1,0,5]$ & $1$ & $12$\\
& $[2,2,0]$ & $2$ & $-$\\
& $[2,0,4]$ & $2$ & $-$\\
& $[4,0,2]$ & $2$ & $-$\\
& $[0,2,2]$ & $2$ & $-$\\
& $[0,3,0]$ & $3$ & $-$\\
& $[3,0,3]$ & $3$ & $-$\\
& $[6,0,0]$ & $6$ & $-$\\
& $[0,0,6]$ & $6$ & $-$\\
\end{tabular}

\caption{ Sets $F_M(C_2)$ for $M=1,2,3,6$. Each point of $F_M(C_2)$ is represented by the coordinates $[s_0,s_1,s_2]$ from~(\ref{FM}). The third column contains the full order $k$ of the corresponding conjugacy class of the elements of finite order.}\label{efo}
\end{table}
}

\subsection{Interior of $F_M$}\

Subsequently we will need to know the order of the stabilizer of a given point in the grid $F_M$. The stabilizer is trivial if the element is found in the interior of $F_M$. Below we consider special functions called $S-$functions and their discrete pairwise orthogonality. Only the points from the interior of the $F_M$ are involved in the orthogonality relations of $S-$functions.
Define an interior of the grid $F_M$
as such cosets from $\frac{1}{M}P^{\vee}/Q^{\vee}$ which have representative element in the interior of $F$:
\begin{equation*}
\wt F_M\equiv\frac{1}{M}P^{\vee}/Q^{\vee}\cap \mathrm{int}( F).
 \end{equation*}

Points from $F_M$ which have all coordinates $s_0,s_1,\dots,s_n$ positive are precisely those from $\wt F_M$. Thus, from (\ref{FM}) we see that the set $\wt F_M$ can be identified as
\begin{equation}\label{FMint}
 \wt F_M = \setb{\frac{s'_1}{M}\om^{\vee}_1+\dots+\frac{s'_n}{M}\om^{\vee}_n}{s'_0,s'_1,\dots ,s'_n \in \N,\, s'_0+s'_1m_1+\dots + s'_n m_n=M}.
 \end{equation}
We calculate the number of elements of $\wt F_M$.
\begin{tvr}\label{interior}
Let $m$ be the Coxeter number. Then
$$|\wt F_M|=\begin{cases}0 & M<m \\ 1 & M=m \\ |F_{M-m}| & M>m .\end{cases}$$
\end{tvr}
\begin{proof}
Substituing the relations $s'_i=1+s_i, s_i\in \Z^{\geq 0},i\in\{0,\dots,n\}$ into the defining relation (\ref{FMint}) we obtain
\begin{equation*}
 s_0+m_1s_1+\dots + m_n s_n= M-m,\q s_0,\dots,s_n\in \Z^{\geq 0}.
\end{equation*}
This equation has one solution $[0,\dots,0]$ if $M=m$, no solution if $M<m$ and is equal to the defining relation (\ref{FM}) of $F_{M-m}$ if $M>m$.
\end{proof}
\begin{example}
For the algebra $E_8$ we have the Coxeter number $m=30$. We calculated in Example \ref{E8num} that $|F_{30}(E_8)|=20956$. According to Proposition \ref{interior}, among these $20956$ points is only one in the interior of $F_{30}(E_8)$, i. e. $|\wt F_{30}(E_8)|=1$.
\end{example}

\begin{example}\label{ex1}
For the Lie algebra $C_2$ we have Coxeter number $m=4$ and $c=2$. Consider for example $M=4$. For the order of the group $\frac{1}{4}P^{\vee}/Q^{\vee}$ we have from (\ref{grupa}) that $\abs{\frac{1}{4}P^{\vee}/Q^{\vee}}=32$ and according to Theorem \ref{numAn} we calculate $$\abs{F_4(C_2)}=\comb{4}{2}+\comb{3}{2}=9. $$
Note also that from Proposition \ref{interior} follows the number of points in the interior $|\wt F_4(C_2)|=1$. The cosets representants of $\frac{1}{4}P^{\vee}/Q^{\vee}$ and fundamental domain $F$ are depicted in Figure \ref{figC2}.
\begin{figure}
\resizebox{11cm}{!}{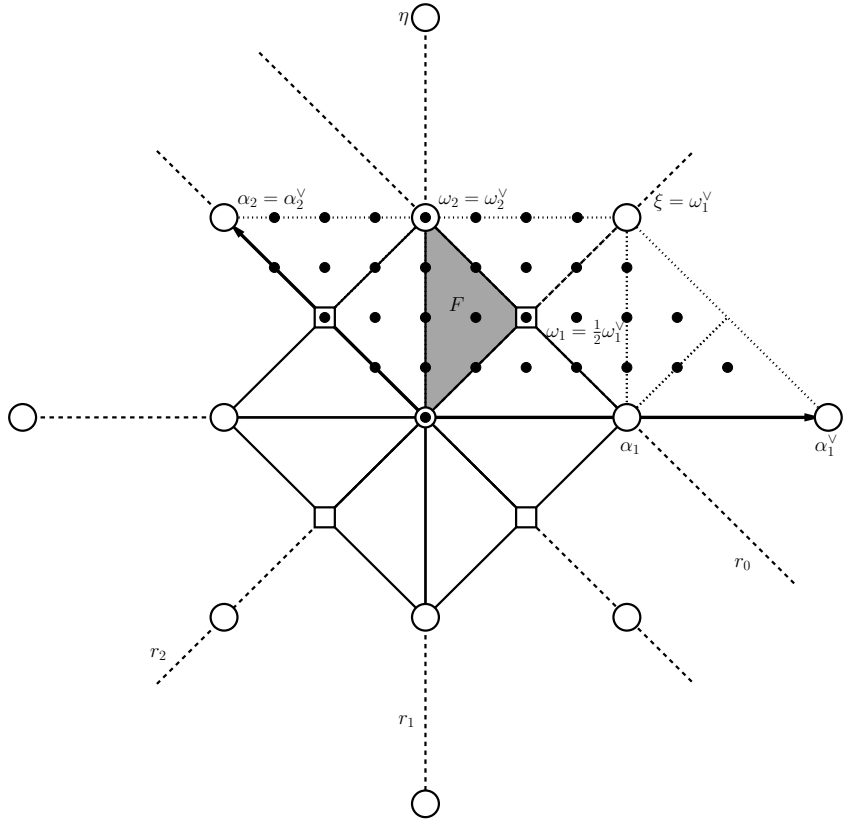}
\caption{ The cosets representants of $\frac{1}{4}P^{\vee}/Q^{\vee}$ of $C_2$; the cosets representants are shown as $32$ black dots, the grey area is the fundamental domain $F$ which contains $9$ points of $F_4(C_2)$. The dashed lines represent 'mirrors' $r_0,r_1$ and $r_2$. Circles are elements of root lattice $Q$, together with squares they are elements of the weight lattice $P$.  }\label{figC2}
\end{figure}
\end{example}

\subsection{Grid $\Lambda_M$}\

The points of $\Lambda_M$ are the dominant weights specifying $C$ or $S-$functions which belong to the same pairwise orthogonal set. Later on we consider $C-$ or $S-$functions which are sampled on the points $F_M$. We consider the lowest possible set of such points. The number of points of $\Lambda_M$ coincides with the number of points of $F_M$.

The $W$-invariant group $P/MQ$ is isomorphic to the group $\frac{1}{M}P^{\vee}/Q^{\vee}$, thus its order is given by the formula
\begin{equation*}
\abs{P/MQ}=cM^n.
\end{equation*}
Define the grid $\Lambda_M$ as such cosets from $P/MQ$ which has some representative element in~$M F^\vee$
\begin{equation*}
 \Lambda_M\equiv M F^\vee \cap P/MQ.
 \end{equation*}

We have from (\ref{defdfun}) that the set $\Lambda_M$, or more precisely its representative points, can be identified as
\begin{equation}\label{LAM}
\Lambda_M = \setb{t_1\om_1+\dots+t_n\om_n}{t_0,t_1,\dots ,t_n \in \Z^{\geq 0},\, t_0+t_1m^{\vee}_1+\dots + t_n m^{\vee}_n=M}.
 \end{equation}

Observing Table~\ref{tabmarks}, we see that the $n$--tuple of dual marks $(m^{\vee}_1,\dots,m^{\vee}_n)$ is a certain permutation of the $n$--tuple $(m_1,\dots , m_n)$. Thus, we have from (\ref{FM}) and (\ref{LAM}) that
\begin{equation}\label{FL}
 |F_M|=|\Lambda_M|.
\end{equation}

If we have two elements $b,b'\in \R^n$ such that $b'-b=Mq$, with $q\in Q$ then for $w\in W$ we have $wb-wb'=wMq\in MQ$, i.e. we have a natural action of $W$ on the quotient group $\R^n/MQ$. For $\la \in \R^n/MQ$ we denote
 \begin{equation*}
\mathrm{Stab}^{\vee} (\la)=\set{w\in W}{w\la=\la}
\end{equation*}
and
\begin{equation}\label{hla}
h^{\vee}_{\la}\equiv |\mathrm{Stab}^{\vee} (\la)|.
\end{equation}
\begin{tvr}
\begin{enumerate}
\item For any $\la\in P/MQ$ there exists $\la'\in\Lambda_M  $ and $w\in W$ such that
\begin{equation}\label{dfun1}
 \la=w\la'.
\end{equation}
\item If $\la,\la'\in \Lambda_M $ and $\la'=w\la$, $w\in W$ then
\begin{equation}\label{lrfun2}
 \la'=\la=w\la.
\end{equation}

\item If $\la\in M F^\vee \cap \R^n/MQ $, i.e. $\la=b+MQ$, $b\in MF^\vee$ then
\begin{equation}\label{rfunstab2}
\mathrm{Stab}^{\vee} (\la) \cong \mathrm{Stab}_{\widehat{W}^{\mathrm{aff}}}(b/M).
\end{equation}
\end{enumerate}
\end{tvr}
\begin{proof}
\begin{enumerate}
 \item Let $\la\in P/MQ$ be of the form $\la=p+MQ $, $p\in P$. From (\ref{ddfun1}) follows that there exist $p'\in F^\vee$, $w\in W$ and $q\in Q$ such that
\begin{equation*}\label{ddfun2}
 \frac1M p=w p'+q,
 \end{equation*}
i.e. $ p=wMp'+Mq$.
From $W$--invariance of $P$ we have that $Mp'\in P$, the class $\la'=Mp'+MQ$ is from $\Lambda_M$
and (\ref{dfun1}) holds.
\item Let $\la,\la'\in P/MQ$ be of the form $ \la=p+MQ$, $\la'=p'+MQ$ and $p,p'\in MF^\vee$. Suppose that
\begin{equation*}\label{ddfun4}
 p'=wp+Mq,\q q\in Q,\, w\in W.
\end{equation*}
Then $p/M,p'/M\in F^\vee$ and it follows from (\ref{dfun2}) that $p=p'$.
\item We have from (\ref{directd}) that for any $w^{\mathrm{aff}}\in \widehat{W}^{\mathrm{aff}}$ there exist unique $w\in W$ and unique shift $T(q)$ such that $w^{\mathrm{aff}}=T(q)w$. Define map $\psi: \mathrm{Stab}_{\widehat{W}^{\mathrm{aff}}}(b/M)\map W $ for $w^{\mathrm{aff}}\in \mathrm{Stab}_{\widehat{W}^{\mathrm{aff}}}(b/M)$ by
\begin{equation*}\label{izo11d}
\psi(w^{\mathrm{aff}})=\psi(T(q)w)=w.
\end{equation*}
Since the relation
\begin{equation*}\label{izo2d}
\psi(T(q_1)w_1T(q_2)w_2)=\psi(T(q_1+w_1q_2)w_1w_2)=w_1w_2
\end{equation*}
holds, $\psi$ is indeed a homomorphism. If $b/M=w^{\mathrm{aff}}(b/M)=w(b/M)+q$ then $b-wb=Mq\in MQ$, i.e. we obtain $w\in \mathrm{Stab}^{\vee} (\la)$ and vice versa. Thus, $\psi (\mathrm{Stab}_{\widehat{W}^{\mathrm{aff}}}(b/M))=\mathrm{Stab}^{\vee}(\la)$ holds. We also have
\begin{equation*}\label{kerd}
\mathrm{ker}\,\psi= \left\{ T(q)\in \mathrm{Stab}_{\widehat{W}^{\mathrm{aff}}}(b/M) \right\}=1.
\end{equation*}
\end{enumerate}
\end{proof}

\subsection{Interior of $\Lambda_M$}\

The points of the interior of $\Lambda_M$ are the points which label non-zero $S-$functions. For the points on the boundary of $\Lambda_M$, the $S-$functions are equal to zero.
Define an interior of the grid $\Lambda_M$
as such cosets from $P/MQ$ which have representative element in the interior of $M F^\vee$
\begin{equation*}\label{LMcint}
\wt \Lambda_M\equiv P/MQ\cap \mathrm{int}(M F^\vee).
 \end{equation*}

Analogously to (\ref{FMint}), the set $\wt \Lambda_M$ can be identified as
\begin{equation}\label{LMint}
\wt \Lambda_M = \setb{t'_1\om_1+\dots+t'_n\om_n}{t'_0,t'_1,\dots ,t'_n \in \N,\, t'_0+t'_1m^{\vee}_1+\dots + t'_n m^{\vee}_n=M}
 \end{equation}
and moreover
\begin{equation}\label{FLint}
 |\wt F_M|=|\wt \Lambda_M|.
\end{equation}

\begin{example}\label{exdual}
For the Lie algebra $C_2$ we have  $\abs{P/4Q}=32$ and according to (\ref{FL}) we have $$\abs{\Lambda_4(C_2)}=\abs{F_4(C_2)}=9. $$
Note also that from (\ref{FLint}) follows the number of points in the interior $|\wt \Lambda_4(C_2)|=1$. The cosets representants of $P/4Q$, the dual fundamental domain $F^{\vee}$ and the grid $\Lambda_4(C_2)=4F^{\vee}\cap P/4Q$ are depicted in Figure~\ref{figC2d}.
\begin{figure}
\resizebox{11cm}{!}{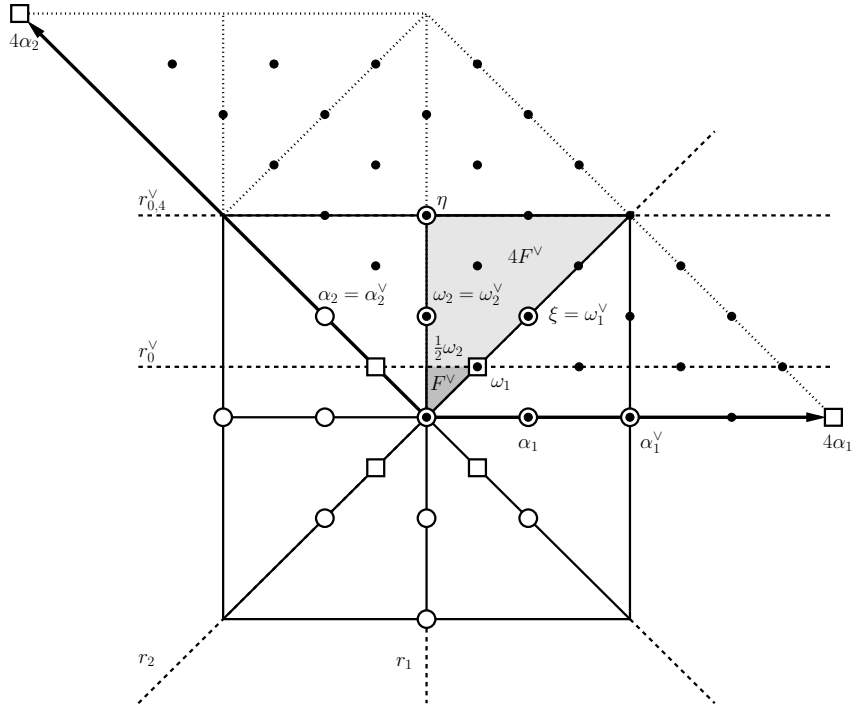}
\caption{ The cosets representants of $P/4Q$ of $C_2$; the cosets representants are shown as $32$ black dots, the darker grey area is the fundamental domain $F^{\vee}$, the lighter grey area is the domain $4F^{\vee}$ which contains $9$ elements of $\Lambda_4(C_2)$. The dashed lines represent dual 'mirrors' $r^\vee_0,r_1$, $r_2$ and the affine mirror $r^\vee_{0,4}$ (see~(\ref{affmir})). The circles and squares coincide with those in Figure \ref{figC2}. }\label{figC2d}
\end{figure}
\end{example}

\subsection{Calculation of $h_x$ and $h^{\vee}_\la$}\

Calculation procedure of $h_x$ for any $x\in F_M$ can be deduced from~(\ref{stab}) and~(\ref{rfunstab}). It uses the extended $\mathrm{DD}$ of~$G$. Using the extended $\mathrm{DD}^{\vee}$ of~$G$, we obtain from~(\ref{stabdual}) and~(\ref{rfunstab2}) analogous calculation procedure for $h^{\vee}_\la$, $\la\in \Lambda_M$.

Consider a point $x\in F_M$ and its corresponding coordinates $[s_0,\dots,s_n]$ from (\ref{FM}).
\begin{enumerate}
 \item If $s_0,\dots,s_n$ are all non-zero then $h_x=1$.
 \item Consider such a subgraph $U$ of extended $\mathrm{DD}$ consisting only of those nodes $i$ for which $s_i=0,\, i=0,\dots,n$. The subgraph $U$ consists in general of several connected components $U_l$. Each component~$U_l$ is a (non-extended) $\mathrm{DD}$ of some simple Lie group~$G_l$. Take corresponding orders of the Weyl groups $|W_l|$ of $G_l$ from Table \ref{tab1}. Then $h_x=\prod_{l}|W_l|$.
\end{enumerate}

 We proceed similarly to determine $h^{\vee}_\la$ when we consider a point $\lambda\in \Lambda_M$ and its corresponding coordinates $[t_0,\dots,t_n]$ from (\ref{LAM}).
\begin{enumerate}
 \item If $t_0,\dots,t_n$ are all non-zero then $h^{\vee}_\la=1$.
 \item Consider such a subgraph $U'$ of the extended $\mathrm{DD}^{\vee}$ consisting only of those nodes $i$ for which $t_i=0,\, i=0,\dots,n$. The subgraph $U'$ consists in general of several connected components $U'_l$. Each component~$U'_l$ is a (non-extended) $\mathrm{DD}$ of some simple Lie group~$G'_l$. Take corresponding orders of the Weyl groups $|W'_l|$ of $G'_l$ from Table \ref{tab1}. Then $h^{\vee}_\la=\prod_{l}|W'_l|$.
\end{enumerate}

\section{$W-$Invariant functions}
The numbers $h_x$, $h^{\vee}_\la$ and $|F_M|$, which were determined so far, are important for the properties of special functions called $C-$ and $S-$functions when they sampled on $F_M$. A detailed review of the properties of $C-$ and $S-$functions may be found in \cite{KP1,KP2}. In this section we want to complete and make explicit the orthogonality properties of $C-$ and $S-$functions \cite{MP2}. Two and three-dimensional examples of these relations are found in~\cite{PZ1,PZ2,PZ3,PA,NP}.
\subsection{$C$--functions}\
We recall the definition of $C-$functions and show that they can be labeled by the finite set $\Lambda_M$ when sampled on the grid $F_M$.

Consider $ b\in P$ and recall that (normalized) $C-$functions can be defined as a mapping $\Phi_b:\R^n\map \Com$
\begin{equation}\label{C}
 \Phi_b(a)=\sum_{w\in W} e^{2 \pi \i \sca{ wb}{a}}.
 \end{equation}
The following properties of $C-$functions are crucial
\begin{itemize}
\item symmetry with respect to $w\in W$
\begin{align}\label{Csym}
 \Phi_b(wa)&=\Phi_b(a) \\
\Phi_{wb}(a)&=\Phi_b(a) \label{Csymla}
 \end{align}
\item invariance with respect to $q^{\vee} \in Q^{\vee}$
\begin{equation}\label{Cinv}
 \Phi_b(a+q^{\vee})=\Phi_b(a).
 \end{equation}
\end{itemize}

We investigate the values of $C-$functions on the grid $F_M$. Suppose we have fixed $M\in \N$ and $s\in \frac{1}{M}P^{\vee}$. From (\ref{Cinv}) follows that we can consider $\Phi_b$ as a function on classes $\frac{1}{M}P^{\vee}/Q^{\vee}$. From (\ref{rfun1}) and (\ref{Csym}) follows that we can consider $\Phi_b$ only on the set $F_M\equiv\frac{1}{M}P^{\vee}/Q^{\vee}\cap F$. We also have
\begin{equation*}
 \Phi_{b+MQ}(s)=\Phi_b(s),\ s\in F_M
 \end{equation*}
and thus we can consider the functions $\Phi_\la$ on $F_M$ parameterized by classes from $\la\in P/MQ$. Moreover, from (\ref{dfun1}) and (\ref{Csymla}) follows that {\it we can consider $C-$functions $\Phi_\la$ on $F_M$ parameterized by $\la\in\Lambda_M$ only}.

\subsection{$S-$functions}\

We recall the definition of $S-$functions and show that they can be labeled by the finite set $\wt\Lambda_M$ when sampled on the grid $\wt F_M$.

Consider $b\in P$ and recall that (normalized) $S-$functions can be defined as a mapping $\phi_b:\R^n\map \Com$
\begin{equation}\label{S}
 \phi_b(a)=\sum_{w\in W}(\det w) e^{2 \pi \i \sca{ wb}{a}}
 \end{equation}
The following properties of $S-$functions are crucial \begin{itemize}
\item antisymmetry with respect to $w\in W$
\begin{align}\label{Ssym}
 \phi_b(wa)&=(\det w) \phi_b(a) \\
\phi_{wb}(a)&=(\det w)\phi_b(a) \label{Ssymla}
 \end{align}
\item invariance with respect to $q^{\vee} \in Q^{\vee}$
\begin{equation}\label{Sinv}
 \phi_b(a+q^{\vee})=\phi_b(a).
 \end{equation}
\end{itemize}
Moreover, simple calculation shows that
$
 \phi_b(r_0 a)=- \phi_b(a)
$
and thus we have
\begin{equation}\label{Sri}
 \phi_b(r_i a)=- \phi_b(a), \q i\in\{0,\dots,n\}.
\end{equation}

We investigate the behavior of $S-$functions on the grid $F_M$. Suppose we have fixed $M\in \N$ and $s\in \frac{1}{M}P^{\vee}$. From (\ref{Sinv}) follows that we can consider $\phi_b$ as a function on classes $\frac{1}{M}P^{\vee}/Q^{\vee}$. From (\ref{rfun1}) and (\ref{Ssym}) follows that we can consider $\phi_b$ only on the set $F_M\equiv\frac{1}{M}P^{\vee}/Q^{\vee}\cap F$. Moreover, from (\ref{Sri}) we see that $\phi_b$ vanishes on the boundary of the fundamental domain $F$ and thus {\it we can consider $\phi_b$ only on the interior $\wt F_M$}.

We also have
\begin{equation*}
 \phi_{b+MQ}(s)=\phi_b(s), \q s\in \wt F_M
 \end{equation*}
and thus we can consider the functions $\phi_\la$ on $\wt F_M$ parameterized by classes from $\la\in P/MQ$. Moreover, from (\ref{dfun1}) and (\ref{Ssymla}) follows that we can consider $\phi_\la$ on $\wt F_M$ parameterized by classes from $\Lambda_M$. From (\ref{Ssymla}) we have that
\begin{equation}\label{Sr0d}
 \phi_{r^{\vee}_i\la}( s)=- \phi_\la(s),\q s\in \wt F_M,\q i\in\{1,\dots,n\}.
 \end{equation}

For affine reflection $r_{0,M}^{\vee}$, given by
\begin{equation}\label{affmir}
r_{0,M}^{\vee} \la=r_{\eta} \la + M\frac{2\eta}{\sca{\eta}{\eta}}
\end{equation}
we also obtain
\begin{equation}\label{Sr0aff}
 \phi_{r_{0,M}^{\vee}\la}( s)=- \phi_\la(s),\q s\in \wt F_M.
 \end{equation}
We conclude that for $\la\in\Lambda_M\setminus \wt\Lambda_M$
$$
 \phi_{\la}( s)=0,\q s\in \wt F_M.
$$
Thus, {\it we can consider $S-$functions $\phi_\la$ on $\wt F_M$ parameterized by $\la\in\wt \Lambda_M$ only}.

\section{Discrete orthogonality of $C-$ and $S-$functions}

\subsection{Basic discrete orthogonality relations}\

Discrete orthogonality of the $C-$ and $S-$functions was discussed in general in \cite{MP2}. Practical use of \cite{MP2} is not completely straightforward. Therefore we reformulate the basic facts and we subsequently use them to make the discrete orthogonality over $F_M$ described in all details.

Note that since for $a\in\frac{1}{M}P^{\vee}$ and $b\in P $ the relation
$$e^{2\pi\i\sca{b+MQ}{a+Q^{\vee}}}= e^{2\pi\i\sca{b}{a}}$$
holds, the exponential mapping $e^{2\pi\i\sca{\la}{x}}\in\Com$ for $\la \in P/MQ$ and $x\in\frac{1}{M}P^{\vee}/Q^{\vee}$ is well-defined.

\begin{tvr}
For all $\la \in P/MQ$, $\la\neq 0$ there exists $x\in\frac{1}{M}P^{\vee}/Q^{\vee}$ such that $e^{2\pi\i\sca{\la}{x}}\neq 1$.
\end{tvr}
\begin{proof}
Suppose there is some $p\in P$ and $p\notin MQ$ such that for all $p^{\vee}\in P^{\vee}$
\begin{equation}
 \sca{p}{\frac{p^{\vee}}{M}}=\sca{\frac{p}{M}}{p^{\vee}}\in \Z .
\end{equation}
Then from (\ref{dPPP}) we have $p/M\in Q$, i.e. $p\in MQ$ -- a contradiction.
\end{proof}

\begin{cor}
For any $\la,\la' \in P/MQ$ it holds:
\begin{equation}
 \sum_{y\in\frac{1}{M}P^{\vee}/Q^{\vee}} e^{2\pi\i\sca{\la-\la'}{y}}=cM^n\delta_{\la,\la'}.
\end{equation}
 \end{cor}

\subsection{Discrete orthogonality of $C-$functions}\

We define the scalar product of two functions $f,g:F_M\map \Com$ by
\begin{equation}
 \sca{f}{g}_{F_M}= \sum_{x\in F_M}\ep(x) f(x)\overline{g(x)},
\end{equation}
where the numbers $\ep (x)$ are determined by (\ref{ep}). We show that $\Lambda_M$, defined by (\ref{LAM}), is the lowest maximal set of pairwise orthogonal $C-$functions.
\begin{tvr}
For $\la,\la' \in\Lambda_M$ it holds
\begin{equation}\label{ortho}
 \sca{\Phi_\la}{\Phi_{\la'}}_{F_M}=c\abs{W}M^n h^{\vee}_\la \delta_{\la,\la'},
\end{equation}
where $c$, $h^{\vee}_\la$, $\Phi_\la$ were defined by (\ref{Center}), (\ref{hla}), (\ref{C}) respectively, $n$ is the rank of $G$.
\end{tvr}
\begin{proof}
The equality
\begin{equation*}
\sum_{x\in F_M} \ep(x) \Phi_\la(x)\overline{\Phi_{\la'}(x)}= \sum_{y \in \frac{1}{M}P^{\vee}/Q^{\vee}} \Phi_\la(y)\overline{\Phi_{\la'}(y)}
\end{equation*}
follows from (\ref{rfun2}) and (\ref{WFM}) and $W-$invariance of the expression $\Phi_\la(x)\overline{\Phi_{\la'}(x)}$. Then, using $W-$invariance of $\frac{1}{M}P^{\vee}/Q^{\vee}$,  we have
\begin{align*}
\sca{\Phi_\la}{\Phi_{\la'}}_{F_M}=&\sum_{w'\in W}\sum_{w\in W} \sum_{y\in \frac{1}{M}P^{\vee}/Q^{\vee}}e^{2\pi\i\sca{w\la-w'\la'}{y}}=\abs{W}\sum_{w'\in W}\sum_{y \in \frac{1}{M}P^{\vee}/Q^{\vee}}e^{2\pi\i\sca{\la-w'\la'}{y}}\\ =& c\abs{W}M^n \sum_{w'\in W} \delta_{w'\la',\la}.
\end{align*}
Since $\la,\la' \in\Lambda_M$ we have from (\ref{lrfun2}) that $$ \sum_{w'\in W} \delta_{w'\la',\la}=h^{\vee}_{\la}\delta_{\la,\la'}. $$
\end{proof}

\subsection{Discrete orthogonality of $S-$functions}\

Unlike the $C-$functions, the $S-$functions have non-zero values on the interior $\wt F_M$ only. Note that from~(\ref{stab}) and~(\ref{rfunstab}) we have that $\ep (x)=\abs{W}$, $x\in\wt F_M$. Therefore, we define
the scalar product of two functions $f,g:\wt F_M\map \Com$ as follows
\begin{equation}
 \sca{f}{g}_{\wt F_M}=\abs{W} \sum_{x\in \wt F_M} f(x)\overline{g(x)}.
\end{equation}
Then since $\phi_\la$ vanishes on the boundary of $F_M$, we have
$$\sca{\phi_{\la}}{\phi_{\la'}}_{ F_M}=\abs{W} \sum_{x\in \wt F_M} \phi_{\la}(x)\overline{\phi_{\la'}(x)}=\sca{\phi_{\la}}{\phi_{\la'}}_{\wt F_M}. $$
We show that $\wt\Lambda_M$, determined by (\ref{LMint}), is the lowest maximal set of pairwise orthogonal $S-$functions.
\begin{tvr}
 For $\la,\la' \in\wt \Lambda_M$ it holds
\begin{equation}\label{Sortho}
 \sca{\phi_\la}{\phi_{\la'}}_{\wt F_M}=c\abs{W}M^n \delta_{\la,\la'}.
\end{equation}
where $c$, $\phi_\la$ were defined by (\ref{Center}), (\ref{S})  respectively, $n$ is the rank of $G$.
\end{tvr}
\begin{proof}
The equality
\begin{equation*}
\sum_{x\in F_M} \ep(x) \phi_\la(x)\overline{\phi_{\la'}(x)}= \sum_{y \in \frac{1}{M}P^{\vee}/Q^{\vee}} \phi_\la(y)\overline{\phi_{\la'}(y)}
\end{equation*}
follows from (\ref{rfun2}) and (\ref{WFM}) and $W-$invariance of the expression $\phi_\la(x)\overline{\phi_{\la'}(x)}$. Then, using $W-$invariance of $\frac{1}{M}P^{\vee}/Q^{\vee}$,  we have
\begin{align*}
\sca{\phi_\la}{\phi_{\la'}}_{F_M}=&\sum_{w'\in W}\sum_{w\in W} \sum_{y\in \frac{1}{M}P^{\vee}/Q^{\vee}}(\det ww')e^{2\pi\i\sca{w\la-w'\la'}{y}}=\abs{W}\sum_{w'\in W}\sum_{y \in \frac{1}{M}P^{\vee}/Q^{\vee}}(\det w')e^{2\pi\i\sca{\la-w'\la'}{y}}\\ =& c\abs{W}M^n \sum_{w'\in W}(\det w') \delta_{w'\la',\la}.
\end{align*}
If $\la=w'\la'$  then we have from (\ref{lrfun2}) that $w'\la=\la=\la'$. Since $\la \in\wt \Lambda_M$, it follows from~(\ref{stabdual}) and~(\ref{rfunstab2}) that  $w'=1$.
\end{proof}





\subsection{Discrete $C-$ and $S-$transforms}\

Analogously to ordinary Fourier analysis, we define interpolating functions $ \Phi^M, \phi^M$
\begin{align}
\Phi^M(x):=& \sum_{\la\in \Lambda_M} c_\la \Phi_\la(x),\q x\in \R^n \label{intc} \\
\phi^M(x):=& \sum_{\la\in \wt\Lambda_M} \wt c_\la \phi_\la(x), \q x\in \R^n\label{ints}
\end{align}
which are given in terms of expansion functions $\Phi_\la, \phi_\la$ and expansion coefficients $c_\la$, whose values need to be determined. These interpolating functions can also be understood as finite cut-offs of infinite expansions. The interpolation properties of (\ref{intc}), (\ref{ints}) were tested in dimensions two \cite{PZ1,PZ2,PZ3,PA} and three \cite{NP}. In this article we develop the possibility for solving the interpolation problem for general class of functions sampled on $F_M$.

Next we discretize the equations (\ref{intc}),(\ref{ints}). Suppose we have some function $f$ sampled on the grid $F_M$ or $\wt F_M$. The interpolation of $f$ consists in finding the coefficients $c_\la$ (or $\wt c_\la$) in the interpolating functions (\ref{intc}) or (\ref{ints}) such that
\begin{align}
\Phi^M(x)=& f(x), \q x\in F_M \label{intcs}\\
\phi^M(x)=& f(x), \q x\in \wt F_M.\label{intss}
\end{align}
Relations (\ref{FL}), (\ref{FLint}) and (\ref{ortho}), (\ref{Sortho}) allow us to view the values $\Phi_\la(x)$ with $x\in F_M$, $\la\in \Lambda_M$ and the values of $\phi_\la(x)$ with $x\in \wt F_M$, $\la\in \wt\Lambda_M$ as elements of non-singular square matrices.
These invertible matrices coincide with the matrices of the linear systems (\ref{intcs}), (\ref{intss}). Thus, the coefficients $c_\la$, $\wt c_\la$ can be in both cases uniquely determined. The formulas for calculation of $c_\la$ and $\wt c_\la$, which are also called discrete $C-$ and $S-$transforms, can obtained by means of calculation of standard Fourier coefficients:
\begin{align}
c_\la=& \frac{\sca{f}{\Phi_\la}_{F_M}}{\sca{\Phi_\la}{\Phi_\la}_{F_M}}=(c\abs{W} M^nh^{\vee}_\la)^{-1}\sum_{x\in F_M}\ep(x) f(x)\overline{\Phi_\la(x)}\label{ctrans}\\
\wt c_\la=& \frac{\sca{f}{\phi_\la}_{\wt F_M}}{\sca{\phi_\la}{\phi_\la}_{\wt F_M}}=(c M^n)^{-1}\sum_{x\in \wt F_M} f(x)\overline{\phi_\la(x)}.\label{strans}
\end{align}

\section{Concluding Remarks}
\begin{itemize}
\item
In situations when a large number of data (functions), sampled on the same grid $F_M$, has to be developed into finite series, the processing load can be significantly lightened, which may not be evident from the exposition above. The problem can be viewed in the following matrix form.
$$
b=Df\,, \qquad f\in\Com^{|\Lambda_M|\times1}\,,
          \quad b\in\Com^{|F_M|}
         \quad D\in\Com^{|\Lambda_M|\times|F_M|}\,.
$$
Here $f$ is a matrix column containing the values of the function sampled on the points of $F_M$, which is being decomposed into the finite series of $|\Lambda_M|$ terms. $D$ is the decomposition matrix and $b$ is the matrix column formed by the coefficient of the finite series. Crucial observation here is that $D$ depends on the values of the $C-$ or $S-$functions on the points of $F_M$, but {\it not} on the function that is being decomposed. Therefore the matrix $D$ needs to be computed only once and then stored as a look-up table for subsequent computations. Moreover, it was shown in \cite{MP1} that, with some additional constraints, the decomposition matrix  can be made to contain only integer matrix elements.

\smallskip\item
Products of $C-$ and $S-$functions, referring to the same Lie group and the same point $x\in F$, are fully decomposable into the sum of such functions. More precisely,
\begin{align}
 C_\lambda(x)C_{\lambda'}(x)
     &=\sum_kd_kC_k(x)\,,\qquad d_k\in\N\\
 C_\lambda(x)S_{\lambda'}(x)
     &=\sum_kd_kS_k(x)\,,\qquad d_k\in\Z\\
 S_\lambda(x)S_{\lambda'}(x)
     &=\sum_kd_kC_k(x)\,,\qquad d_k\in\Z
\end{align}
The problem here parallels the decomposition of Weyl group orbits which was extensively studied in \cite{HLP}.

\smallskip\item
Reduction of a semisimple Lie group $G$ to its reductive subgroup, say $G'$, implies decomposition of orbit functions of $G$ to the sum of orbit functions of $G'$, so called computation of branching rules. It involves simultaneous transformation of $\lambda$ and $x$ to the corresponding quantities of the subgroup.

\smallskip\item
Another undoubtedly useful property of the orbit functions, which played no role in this paper, is the fact that they are eigenfunctions of the Laplace operator appropriate for the Lie group and that the eigenvalues are known in all cases.

\smallskip\item
There is another type of special functions similar to $C-$ and $S-$functions, so called $E-$functions \cite{P,NP}. Their discretization properties, fundamental domains, orbit sizes and discrete orthogonality relations will be treated in a separated article.

\smallskip\item
The present work raises the question under which conditions converge series of the functions $\{\Phi^M\}_{M=1}^\infty$, $\{\phi^M\}_{M=1}^\infty$ assigned to a function $f:F\map\Com$ by the relations (\ref{intc}), (\ref{ctrans}) and (\ref{ints}), (\ref{strans}).

\smallskip\item The orbit functions are directly related to the symmetric polynomials \cite{Koorn1,Koorn2,Mac, Dunkl}, see also Section 11 in \cite{KP1}. A detailed description of these relations as well as their discretization properties deserve further study.      

\end{itemize}

\section*{Acknowledgements}
The authors are grateful to R.~V.~Moody for fruitful discussions and comments. Work was supported in part by the Natural Sciences and Engineering Research Council of Canada and by the MIND Research Institute, California. J.~H.~is grateful for the postdoctoral fellowship and the hospitality extended to him at the Centre de recherches math\'ematiques, Universit\'e de Montr\'eal.

\section*{Appendix}
We list the explicit form of the matrices $R$ for all simple Lie algebras. These matrices determine the number $|F_M|$ via the relation (\ref{numpoin}).

$A_n,\,n\geq 1$ $$R(A_n)=\begin{pmatrix}1 \end{pmatrix}$$

$B_n,\,n\geq 3$ $$R(B_n)=\begin{pmatrix}1 & 1 \\ 2 & 0  \end{pmatrix}$$

$C_n,\,n\geq 2$
$$R(C_n)=\begin{pmatrix}1 & 1 \\ 2 & 0  \end{pmatrix}$$

$D_n,\,n\geq 4$ $$R(D_n)=\begin{pmatrix}1 &6 & 1 \\ 4 &4 & 0  \end{pmatrix}$$

{\footnotesize

\begin{align*} R(F_4)=&\left(\begin{smallmatrix}1&84&262&84&1\\\noalign{\medskip}1&100&265&66&0\\\noalign{\medskip}3&121&253&55&0\\\noalign{\medskip}4&139&250&39&0\\\noalign{\medskip}8&160&232&32&0\\\noalign{\medskip}10&181&220&21&0\\\noalign{\medskip}17&199&199&17&0\\\noalign{\medskip}21&220&181&10&0\\\noalign{\medskip}32&232&160&8&0\\\noalign{\medskip}39&250&139&4&0\\\noalign{\medskip}55&253&121&3&0\\\noalign{\medskip}66&265&100&1&0\end{smallmatrix}\right)\q \q \q \q \q \q R(G_2)=\left(\begin{smallmatrix}1&4&1\\\noalign{\medskip}1&5&0\\\noalign{\medskip}2&4&0\\\noalign{\medskip}3&3&0\\\noalign{\medskip}4&2&0\\\noalign{\medskip}5&1&0\end{smallmatrix}\right)\end{align*}
\begin{align*}
R(E_6)=&\left(\begin{smallmatrix}1&132&839&839&132&1\\\noalign{\medskip}3&210&945&708&78&0\\\noalign{\medskip}9&309&1017&567&42&0\\\noalign{\medskip}20&433&1038&433&20&0\\\noalign{\medskip}42&567&1017&309&9&0\\\noalign{\medskip}78&708&945&210&3&0\end{smallmatrix}\right) \q \q R(E_7)=\left(\begin{smallmatrix}1&1011&19029&58150&40371&5775&79\\\noalign{\medskip}2&1414&22208&59612&36626&4510&44\\\noalign{\medskip}6&1944&25585&60536&32856&3464&25\\\noalign{\medskip}12&2616&29172&60816&29172&2616&12\\\noalign{\medskip}25&3464&32856&60536&25585&1944&6\\\noalign{\medskip}44&4510&36626&59612&22208&1414&2\\\noalign{\medskip}79&5775&40371&58150&19029&1011&1\\\noalign{\medskip}128&7296&44032&56128&16128&704&0\\\noalign{\medskip}208&9071&47520&53654&13480&483&0\\\noalign{\medskip}318&11142&50748&50748&11142&318&0\\\noalign{\medskip}483&13480&53654&47520&9071&208&0\\\noalign{\medskip}704&16128&56128&44032&7296&128&0\end{smallmatrix}\right)
\end{align*}

$$R(E_8)=\left(\begin{smallmatrix}1&1366597&143626378&1624457989&4240905745&3094001119&595138048&20483167&20956\\\noalign{\medskip}1&1520913&151753498&1668997388&4262344515&3044716089&571669092&18981162&17342\\\noalign{\medskip}3&1691201&160250557&1714067554&4282411955&2995120797&548866679&17576888&14366\\\noalign{\medskip}5&1876896&169126731&1759671908&4301058735&2945259720&526733429&16260804&11772\\\noalign{\medskip}10&2081203&178395851&1805768016&4318297175&2895163955&505251617&15032514&9659\\\noalign{\medskip}15&2303555&188067805&1852357250&4334079150&2844879535&484421845&13883020&7825\\\noalign{\medskip}27&2547531&198155859&1899395763&4348420650&2794434993&464226699&12812121&6357\\\noalign{\medskip}39&2812537&208670852&1946882187&4361275710&2743877271&444664908&11811413&5083\\\noalign{\medskip}63&3102551&219625367&1994771439&4372663930&2693232237&425719479&10880853&4081\\\noalign{\medskip}90&3416994&231030959&2043059877&4382541090&2642547860&407387331&10012581&3218\\\noalign{\medskip}135&3760240&242899625&2091700825&4390930975&2591846890&389652035&9206725&2550\\\noalign{\medskip}187&4131732&255243637&2140688264&4397791365&2541178068&372508803&8455968&1976\\\noalign{\medskip}270&4536288&268074057&2189974725&4403149365&2490561450&355941747&7760553&1545\\\noalign{\medskip}364&4973372&281404062&2239551147&4406965785&2440045746&339944648&7113703&1173\\\noalign{\medskip}505&5448265&295243652&2289369513&4409270855&2389648355&324502190&6515763&902\\\noalign{\medskip}670&5960475&309606700&2339418205&4410027900&2339418205&309606700&5960475&670\\\noalign{\medskip}902&6515763&324502190&2389648355&4409270855&2289369513&295243652&5448265&505\\\noalign{\medskip}1173&7113703&339944648&2440045746&4406965785&2239551147&281404062&4973372&364\\\noalign{\medskip}1545&7760553&355941747&2490561450&4403149365&2189974725&268074057&4536288&270\\\noalign{\medskip}1976&8455968&372508803&2541178068&4397791365&2140688264&255243637&4131732&187\\\noalign{\medskip}2550&9206725&389652035&2591846890&4390930975&2091700825&242899625&3760240&135\\\noalign{\medskip}3218&10012581&407387331&2642547860&4382541090&2043059877&231030959&3416994&90\\\noalign{\medskip}4081&10880853&425719479&2693232237&4372663930&1994771439&219625367&3102551&63\\\noalign{\medskip}5083&11811413&444664908&2743877271&4361275710&1946882187&208670852&2812537&39\\\noalign{\medskip}6357&12812121&464226699&2794434993&4348420650&1899395763&198155859&2547531&27\\\noalign{\medskip}7825&13883020&484421845&2844879535&4334079150&1852357250&188067805&2303555&15\\\noalign{\medskip}9659&15032514&505251617&2895163955&4318297175&1805768016&178395851&2081203&10\\\noalign{\medskip}11772&16260804&526733429&2945259720&4301058735&1759671908&169126731&1876896&5\\\noalign{\medskip}14366&17576888&548866679&2995120797&4282411955&1714067554&160250557&1691201&3\\\noalign{\medskip}17342&18981162&571669092&3044716089&4262344515&1668997388&151753498&1520913&1\\\noalign{\medskip}20956&20483167&595138048&3094001119&4240905745&1624457989&143626378&1366597&1\\\noalign{\medskip}25080&22083567&619291548&3142941905&4218087760&1580489625&135855148&1225367&0\\\noalign{\medskip}30031&23792451&644124828&3191495835&4193940775&1537086969&128431334&1097777&0\\\noalign{\medskip}35667&25610729&669656034&3239626500&4168460735&1494288513&121340556&981266&0\\\noalign{\medskip}42357&27549093&695878165&3287293110&4141699055&1452086633&114575247&876340&0\\\noalign{\medskip}49945&29608720&722809435&3334456900&4113655575&1410517880&108120805&780740&0\\\noalign{\medskip}58881&31800825&750440529&3381079470&4084381995&1369573029&101970315&694956&0\\\noalign{\medskip}68970&34126983&778789517&3427119670&4053882540&1329286075&96109277&616968&0\\\noalign{\medskip}80756&36598901&807844733&3472541685&4022208950&1289646319&90531409&547247&0\\\noalign{\medskip}94028&39218481&837624096&3517302275&3989369340&1250685567&85222184&484029&0\\\noalign{\medskip}109410&41998015&868113375&3561368375&3955415600&1212391365&80176135&427725&0\\\noalign{\medskip}126672&44939778&899330175&3604694885&3920359630&1174793268&75378787&376805&0\\\noalign{\medskip}146559&48056490&931257831&3647251755&3884252805&1137877656&70825245&331659&0\\\noalign{\medskip}168795&51350948&963913357&3688992120&3847111165&1101671300&66501387&290928&0\\\noalign{\medskip}194259&54836274&997277587&3729889215&3808985225&1066159616&62402841&254983&0\\\noalign{\medskip}222655&58515700&1031366970&3769894675&3769894675&1031366970&58515700&222655&0\\\noalign{\medskip}254983&62402841&1066159616&3808985225&3729889215&997277587&54836274&194259&0\\\noalign{\medskip}290928&66501387&1101671300&3847111165&3688992120&963913357&51350948&168795&0\\\noalign{\medskip}331659&70825245&1137877656&3884252805&3647251755&931257831&48056490&146559&0\\\noalign{\medskip}376805&75378787&1174793268&3920359630&3604694885&899330175&44939778&126672&0\\\noalign{\medskip}427725&80176135&1212391365&3955415600&3561368375&868113375&41998015&109410&0\\\noalign{\medskip}484029&85222184&1250685567&3989369340&3517302275&837624096&39218481&94028&0\\\noalign{\medskip}547247&90531409&1289646319&4022208950&3472541685&807844733&36598901&80756&0\\\noalign{\medskip}616968&96109277&1329286075&4053882540&3427119670&778789517&34126983&68970&0\\\noalign{\medskip}694956&101970315&1369573029&4084381995&3381079470&750440529&31800825&58881&0\\\noalign{\medskip}780740&108120805&1410517880&4113655575&3334456900&722809435&29608720&49945&0\\\noalign{\medskip}876340&114575247&1452086633&4141699055&3287293110&695878165&27549093&42357&0\\\noalign{\medskip}981266&121340556&1494288513&4168460735&3239626500&669656034&25610729&35667&0\\\noalign{\medskip}1097777&128431334&1537086969&4193940775&3191495835&644124828&23792451&30031&0\\\noalign{\medskip}1225367&135855148&1580489625&4218087760&3142941905&619291548&22083567&25080&0\end{smallmatrix}\right)$$

\end{document}